\documentclass[aps,pre,twocolumn,groupedaddress,showpacs]{revtex4}

\usepackage{graphicx}

\begin{document}

\title{Phase diagram of a solution undergoing inverse melting}

\author{R.~Angelini$^{1}$, G.~Ruocco$^{1,2}$, S.~De Panfilis$^{1}$ and F.~Sette$^{3}$ }
\affiliation{$^{1}$ Research center SOFT INFM-CNR, Universit\`a di Roma "La Sapienza" I-00185, Roma, Italy.\\
$^{2}$ Dipartimento di Fisica, Universit\`a di Roma "La Sapienza" I-00185, Roma, Italy.\\
$^{3}$ European Synchrotron Radiation Facility B.P. 220 F-38043 Grenoble, Cedex France.\\
}
\begin{abstract}

The phase diagram  of $\alpha$-cyclodextrin/water/4-methylpyridine
solutions, a system undergoing inverse melting, has been studied
by differential scanning calorimetry, rheological methods, and
X-rays diffraction. Two different fluid phases separated by a
solid region have been observed in the high $\alpha$-cyclodextrin
concentration range ($c$$\geq$150 mg/ml). Decreasing $c$, the
temperature interval where the solid phase exists decreases and
eventually disappears, and a first order phase transition is
observed between the two different fluid phases.

\end{abstract}


\maketitle

\date{\today}
{\it Inverse melting} indicates the counter-intuitive phenomenon
of the solidification of a liquid upon heating. This phenomenon
implies the existence of solids with entropy higher than their
liquid counterparts, and this is recently attracting the attention
of both theoretical and experimental works
~\cite{Ras91aNat,Ras99aMac,Gre00aNat,Por03aNat,Sti03aBio,Sch05aPRE,Sel06aPRB,Sch04aPRL,Cri05aPRL,Fee03aJCP}.
A list of materials showing inverse melting are reported in
ref.~\cite{Sch05aPRE}. Theoretical studies, due to the
difficulties of describing real systems, developed different {\it
ad-hoc} models. All these models allow for the competition between
different interparticle multiple interactions, a feature which may
lead to "crystal" with an entropy content higher than the
corresponding "liquid". Examples are: i) a spin model for fragile
glass-forming liquids~\cite{Sel06aPRB}; ii) spin
glasses~\cite{Sch04aPRL,Sch05aPRE,Cri05aPRL}; and iii) particles
interacting with a specific model potential~\cite{Fee03aJCP}.
These studies derive the temperature {\it vs} crystal field phase
diagrams, which turn out to be characterized by: i) an inverse
transition from a paramagnetic phase to a glass~\cite{Sel06aPRB},
or to a spin glass~\cite{Cri05aPRL,Sch05aPRE,Sch04aPRL} phase for
certain values of the crystal field; ii) a reentrance in the
transition line~\cite{Cri05aPRL,Sch05aPRE,Sch04aPRL,Sel06aPRB};
and iii) a first-order transition line between two paramagnetic
phases ending in a critical point at high values of the crystal
field~\cite{Sel06aPRB}.

On the experimental side, real systems showing inverse melting,
also display complex and exotic phase diagrams due to the variety
of microscopic interactions. This, together with the extreme
thermodynamic conditions often required, prevents comprehensive
investigations of the inverse melting phenomenon. With this
respect, a particularly interesting system is the solutions
composed of $\alpha$-cyclodextrin ($\alpha$CD)
(C$_{36}$H$_{60}$O$_{30}$), water and 4-methylpyridyne (4MP)
(C$_6$H$_7$N). This system has been recently discovered by
Plazanet et al.~\cite{Pla04aJCP} to undergo inverse melting at
easily accessible thermodynamic conditions. These solutions, in
fact, show a "low temperature" liquid (LTL) phase which solidifies
upon heating~\cite{Pla04aJCP,Ang07aPM,Tom05aJCP} into a high
temperature crystal phase (HTC). Therefore, they provide an
"ideal" situation to study experimentally the variety of
phenomenologies recently predicted theoretically.

In this work, using differential scanning calorimetry, rheological
methods and X-rays diffraction, we further investigate the phase
diagram of $\alpha$-cyclodextrin ($\alpha$CD)
(C$_{36}$H$_{60}$O$_{30}$), water and 4-methylpyridyne (4MP)
(C$_6$H$_7$N) solutions, extending the concentration and
temperature ranges with respect to previous studies. At $\alpha$CD
concentrations $c$$\geq$150 mg/ml we find that the "high
temperature" crystal (HTC), if further heated, melt into a fluid
phase, that we refer to as the "high temperature" fluid (HTF). By
decreasing the concentration, the temperature range of existence
of the HTC decreases, and eventually disappears, and we observe a
phase transition between two fluid phases. Specifically, we found
that: i) in the range of concentration $c$$\geq$150 mg/ml of
$\alpha$CD in 4MP, two fluid phases exist in two different
temperature ranges, separated by a temperature interval where the
material is in a crystalline phase, ii) at concentration smaller
than 150 mg/ml, the crystal phase is no longer observable and the
two fluids are contiguous in the phase diagram, and iii) the phase
transition between the two fluids is of first order nature.

\begin{figure}
\begin{center}
\includegraphics[width=8cm,height=10.5cm]{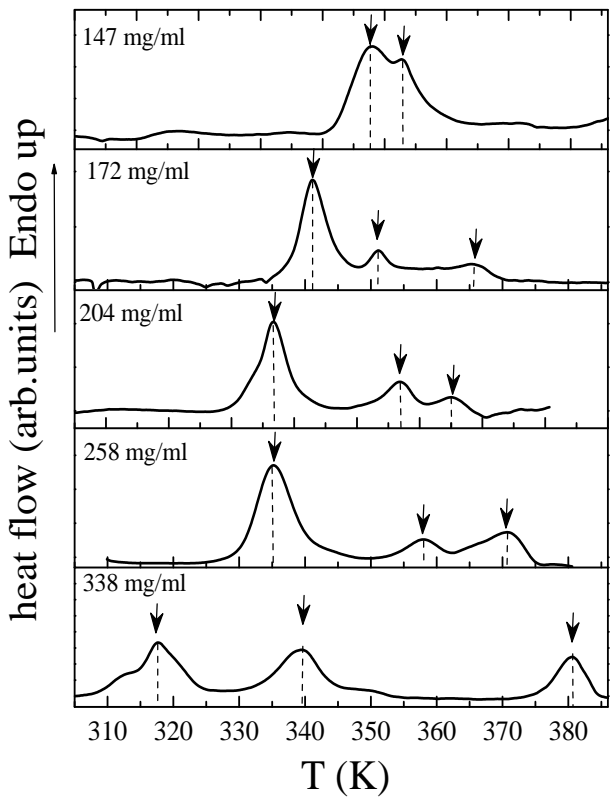}
\caption{ DSC thermograms of solutions of  $\alpha$CD-water-4MP at
the indicated concentrations of $\alpha$CD in 4MP and at a heating
rate r = 10 K/min. The transition peaks are indicated by the
arrows.} \label{fig1}
\end{center}
\end{figure}

Differential scanning calorimetric (DSC) measurements, using a
Diamond Perkin-Elmer calorimeter, have been used to locate the
different phase transitions, and to measure the latent heat of
transformation. Sealed standard aluminium sample pans with a
volume of 10 and 50 $\mu$l have been used as cells and references.
Samples have been prepared using alpha-cyclodextrin hydrate
(Aldrich), water and 4MP (Aldrich) at different concentrations
with molar ratios respectively of 1:6:x, and x varying between
30-200. The powder was dispersed in 4MP and water and then stirred
for  almost 4 hours until the suspensions were cleared. The
thermograms shown in Fig~\ref{fig1}, i.e. the heat flows ($dH/dt$)
as a function of temperature, have been obtained at a heating rate
$r$=10 K/min. The measurements were repeated at least five times
for each concentration, and all transitions have been found to be
reproducible and reversible.

In the concentration region 170-340 mg/ml of $\alpha$CD in 4MP, as
shown in Fig.~1, three endothermic peaks are detected on
increasing temperature: the first one, signature of inverse
melting, corresponds to the LTL-HTC transition \cite{Pla04aJCP},
the intermediate one has been associated to a solid-solid
transition between two of the five crystalline phases detected by
X-rays on similar samples~\cite{Pla06aJCP}. The third one
corresponds to the melting of HTC into the HTF phase.

\begin{figure}[t]
\begin{center}
\includegraphics[width=8.5cm,height=8cm]{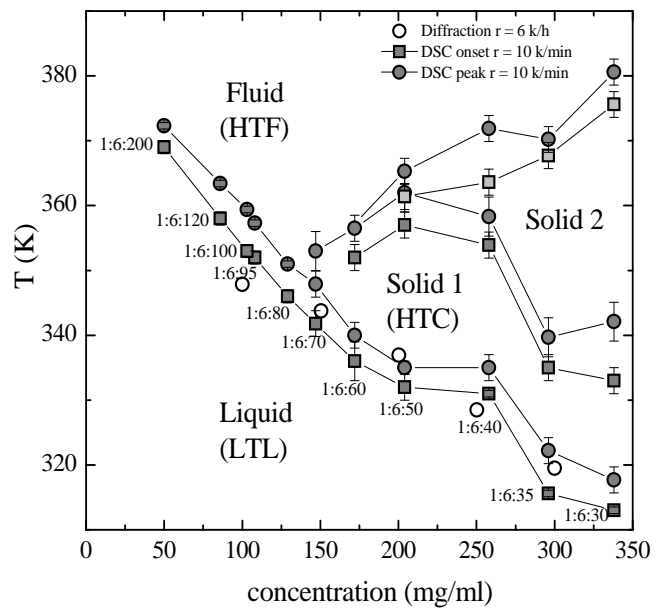}
\caption{Phase diagram of  $\alpha$CD-water-4MP solutions. The
transition temperatures obtained with DSC measurements at a
heating rate r = 10 K/min are plotted as a function of the
concentration of $\alpha$CD in 4MP. The full squares and the full
circles represent the onset and the peak transition temperatures
respectively.  The error bars are calculated through the standard
deviation on five measurements at the same concentration. The
transition temperatures as obtained by X ray diffraction
measurements are also shown: the full triangles and the empty
stars represent respectively the temperatures of the jump in the Q
position of the first maximum and first minimum of the scattered
intensity shown in Fig.~\ref{fig4}. Our DSC and X ray data are
also compared with previous neutron scattering measurements on
similar samples~\cite{Pla04aJCP} (open circles), performed at a
heating rate r = 6 K/h.} \label{fig2}
\end{center}
\end{figure}

\begin{figure}[t]
\begin{center}
\includegraphics[width=8cm,height=8cm]{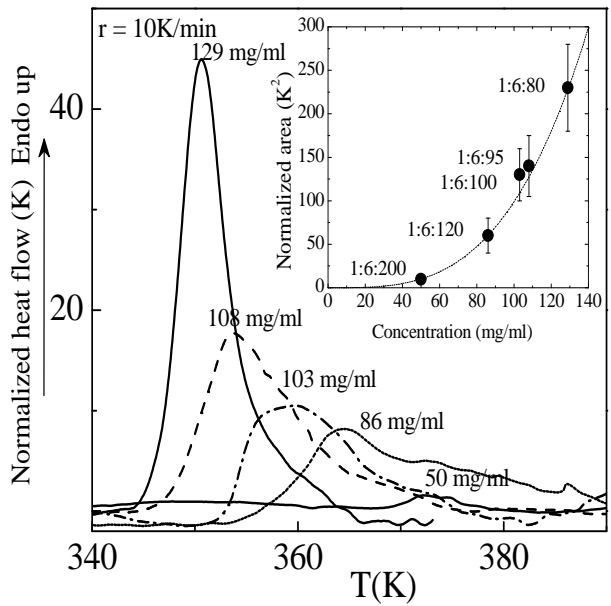}
\caption{Normalized DSC thermograms of  $\alpha$CD-water-4MP
solutions. The measurements have been performed at the indicated
concentrations of $\alpha$CD in 4MP and at a heating rate r = 10
K/min.  A peak of endothermic nature is observed, and is
associated to a phase transition between the LTL and HTF
disordered fluid phases. In the inset, the normalized area of the
peaks is shown as a function of concentration together with a
guideline to the eyes. The error bars represent the standard
deviation on five measurements performed at the same
concentration.}\label{fig3}
\end{center}
\end{figure}

The solid and liquid nature of the HTC, LTL and HTF phases have
been assessed by visual inspection. The transition temperatures,
as determined from the thermograms, are shown as a function of the
concentration in Fig.~\ref{fig2}. The temperatures of the peak
(T$_{peak}$) and of the onset (T$_{onset}$) are shown as full
circles and squares respectively, and provide a range for the
absolute value of the transition temperatures. In the high
concentration range, the phase diagram in Fig.~\ref{fig2}, beside
reproducing the known inverse melting LTL-HTC transition, also
reports the melting of the crystal in the high temperature fluid,
HTF. At lower concentrations, below and close to $c$=150 mg/ml,
not only the intermediate solid-solid transition is no longer
visible, but the whole solid phase disappears and the LTL-HTC and
HTC-HTF transition lines merge into each other. Most importantly,
as shown in Fig.~\ref{fig3}, for concentration $c$$\leq$130 mg/ml,
the thermograms show only one well defined peak.

The endothermic peak of Fig.~\ref{fig3} is observable down to
$c$=50 mg/ml of $\alpha$CD in 4MP, and below this concentration it
falls below the sensitivity of the present experiment. The peak
area is reported in the inset of Fig.~\ref{fig3} as a function of
the $\alpha$CD/4MP mass ratio. These data indicate the decrease of
the heat necessary to the system to complete the LTL-HTF
transition.

\begin{figure}[t]
\begin{center}
\includegraphics[width=8cm,height=9cm]{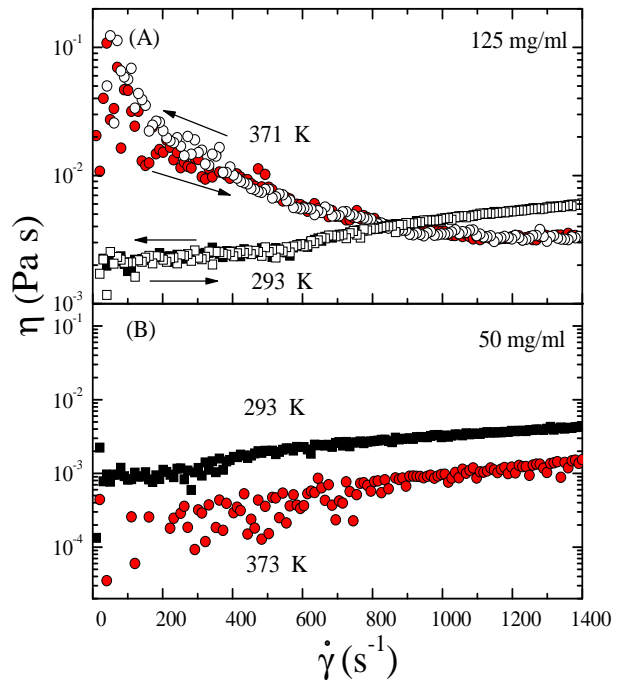}
\caption{(Colors online) Viscosity as a function of the shear rate
at the two indicated concentrations. The samples at c=50 mg/ml
(T=293K, T=373K) and c=125 mg/ml (T=293K) in the LTL phase show
shear thickening while the sample c=125 mg/ml T=371K in the HTF
phase shows shear thinning. As shown in panel (A) the viscosity
profile is perfectly reversible and it can be obtained by
increasing (full symbols) or by decreasing (open symbols) the
applied shear rate.
 } \label{fig4}.
\end{center}
\end{figure}

A different rheological behavior between the LTL and HTF is also
observed in the shear dependent viscosity measurements reported in
Fig.~\ref{fig4}. In Fig.~\ref{fig4}A, as an example, we report
shear viscosity measurements as a function of shear rate at c=125
mg/ml in the HTF (circles) and LTL (squares) phases, which show an
opposite behavior: shear thinning for the HTF and shear thickening
for the LTL. In Fig.~\ref{fig4}B  we report shear viscosity
measurements at almost the same temperatures of Fig.~\ref{fig4}A
for a much lower concentration sample, c=50 mg/ml, in the LTL
phase, the behavior is again shear thickening-like. In conclusion,
the HTF is structurally different from the LTL.

\begin{figure}[t]
\begin{center}
\includegraphics[width=8.5cm,height=8.5cm]{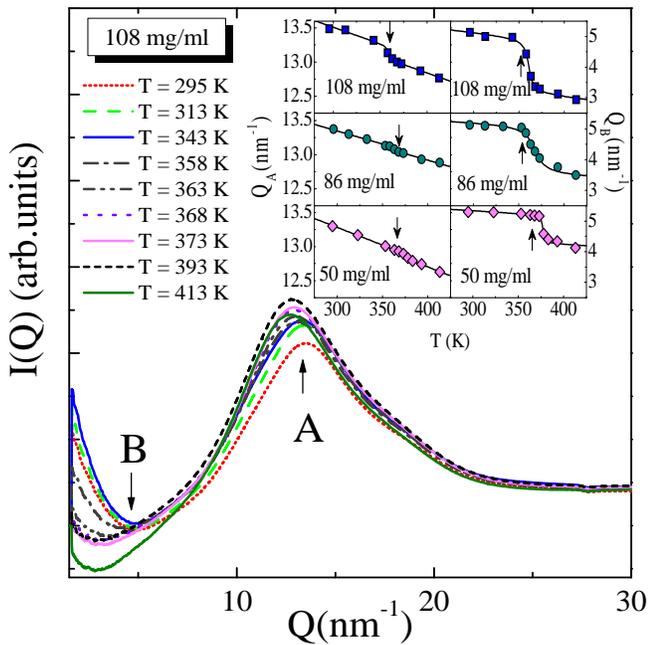}
\caption{(Colors online) Scattered intensity from a solution of
$\alpha$CD, water and 4MP. The measurement  has been done at  the
concentration 108 mg/ml of $\alpha$CD in 4MP as a function of
temperature in the range 295 $\div$ 413 K. In the inset the
Q-position of the maxima Q$_A$ (left panels) and the minima Q$_B$
(right panels) of the scattered intensity are plotted as a
function of temperature for the three indicated concentrations of
$\alpha$CD in 4MP. The arrows indicates the LTL-HTF transition
temperature as derived by DSC. } \label{fig5}
\end{center}
\end{figure}

The phenomenology described so far provides only a macroscopic
description of the two liquid phases, without suggestion on how
they differ from each other at the microscopic level. In order to
investigate this point, the HTF and LTL phases have been studied
by X-ray diffraction. The experiment has been performed on the
beamline BM29 at the European Synchrotron Radiation Facility,
using 15 keV ($\lambda$ = 0.86 $\AA$) incident photons. The
samples were contained in a 2 mm diameter boro-silicate capillary.
The scattered intensity was collected using a MAR345 image plate
detector, and the geometrical parameters were calibrated via Ag
standard. The data analysis was performed using the FIT2D software
package~\cite{Ham96aHig}. We investigated three
$\alpha$CD-water-4MP solutions with $c$=50, 86 and 108 mg/ml. As
an example, data at $c$=108 mg/ml and in the temperature range
295$\div$413 K are shown in Fig.~\ref{fig5} after empty cell
subtraction. The diffraction data, I(Q), reproduce the typical
shape of the static structure factor of a liquid, both below and
above the LTL-HTF transition. More important, the temperature
dependence of I(Q) shows changes in the $Q$-region of the first
minimum and first maximum.

In order to qualitatively represent the temperature changes of the
structural features we arbitrarily selected two representative $Q$
points. Namely, we chose the $Q$ positions of the first maximum
(around $Q$ = 13 nm$^{-1}$) and minimum (around $Q$ = 4
nm$^{-1}$), identified as $Q_A$ and $Q_B$. The $Q$ values of these
two points are reported in the insets of Fig.~\ref{fig5} as a
function of temperature for the three investigated concentrations.
Each curve shows a jump at the temperatures which are reported
(stars) in the phase diagram of Fig.~\ref{fig2}. They, within the
error bars, are in agreement with the DSC measurements. We also
note that the amplitude of the jumps in Fig.~\ref{fig5} are more
pronounced at the higher concentration, suggesting, similarly to
the inset of Fig.~\ref{fig3}, that the phase transition line
between the LTL and HTF tends to disappear on lowering $c$. The
present diffraction data do not allow to determine the structure
of the LTL and HTF phases at the molecular level, although they
clearly underline the presence of a marked structural difference.
These results therefore stimulate further experimental and
theoretical investigations. In particular a point which needs to
be assessed and which cannot be completely excluded  on the basis
of the present measurements, is the presence of a phase separation
in the HTF.

In conclusion, combined DSC, shear viscosity and X ray diffraction
studies on a molecular solution undergoing inverse melting,
allowed us to observe the existence of two fluid phases in two
different temperature ranges separated by a temperature interval
where the material is in a crystalline phase. Most importantly, by
changing the $\alpha$CD concentration control parameter, we
identify the existence of a region in the phase diagram where two
contiguous fluid phases are separated by a first order phase
transition line. These experimental results can be framed within
the recent theoretical  prediction of inverse melting in a spin
model for fragile glasses~\cite{Sel06aPRB}. They also call for
further experimental and simulation works aiming to understand the
interacting mechanism leading to such phenomenology.

R.A. thanks L. Leuzzi for discussions.

\bibliographystyle{apsrev}
%\bibliography{rosa}

\end{document}